\documentclass[aps,prc,twocolumn,nobibnotes,showpacs,superscriptaddress,floatfix]{revtex4-1} 
\usepackage{graphicx} 
\usepackage{dcolumn} 
\usepackage{bm} 

\usepackage{amssymb} 
\usepackage{amsmath} 
\usepackage{color}
\usepackage{xcolor}
\usepackage{hyperref}

\usepackage[caption=false]{subfig}

\begin{document}

\title{Measuring Recoiling Nucleons from the Nucleus with the Electron Ion Collider}

\author{F.~Hauenstein} \affiliation{Old Dominion University, Norfolk, VA}
\affiliation{Massachusetts Institute of Technology, Cambridge, MA}
\author{A.~Jentsch} \affiliation{Brookhaven National Lab, Stony Brook, NY}
\author{J.~R.~Pybus} \affiliation{Massachusetts Institute of Technology, Cambridge, MA}
\author{A.~Kiral} \affiliation{Massachusetts Institute of Technology, Cambridge, MA}
\author{M.~D.~Baker} \affiliation{MDBPADS LLC, Miller Place, NY}
\author{Y.~Furletova} \affiliation{Thomas Jefferson National Accelerator Facility, Newport News, VA}
\author{O.~Hen} \affiliation{Massachusetts Institute of Technology, Cambridge, MA}
\author{D.~W.~Higinbotham} \affiliation{Thomas Jefferson National Accelerator Facility, Newport News, VA}
\author{C.~Hyde} \affiliation{Old Dominion University, Norfolk, VA}
\author{V.~Morozov} \affiliation{Thomas Jefferson National Accelerator Facility, Newport News, VA}
\author{D.~Romanov} \affiliation{Thomas Jefferson National Accelerator Facility, Newport News, VA}
\author{L.B.~Weinstein} \affiliation{Old Dominion University, Norfolk, VA}

\date{\today}

\begin{abstract}
Short range correlated nucleon-nucleon ($NN$) pairs are an important part of the nuclear ground state.  They are typically studied by scattering an electron from one nucleon in the pair and detecting its spectator correlated partner ("spectator-nucleon tagging").   The Electron Ion Collider (EIC) should be able to detect these  nucleons, since they are boosted to high momentum in the lab frame by the momentum of the ion beam.   To determine the feasibility of these studies with the planned EIC detector configuration, we have simulated quasi-elastic scattering for two electron and ion beam energy configurations: 5 GeV $e^{-}$ and 41 GeV/A  ions, and  10 GeV $e^{-}$ and 110 GeV/A ions.  We show that the knocked-out and recoiling nucleons can  be detected over a wide range of initial nucleon momenta.  We also show that  these measurements can achieve much larger momentum transfers than  current fixed target  experiments.  By detecting both low and high initial-momentum   nucleons, the EIC will provide the data that should allow scientists to definitively show if the EMC effect and short-range correlation are connected, and to improve our understanding of color transparency. 
\end{abstract}

\maketitle

\newpage
\section{Introduction}

Short range correlated (SRC) nucleon-nucleon ($NN$) pairs are an important part of the nuclear ground state.  They are typically measured using $(e,e'pN)$ reactions, where the electron scatters from one nucleon and its partner spectator nucleon recoils out of the nucleus.  They account for about 20\% of all nucleons and the vast majority of high momentum nucleons (with momentum greater about 300 MeV/c)~\cite{Subedi:2008zz,LabHallA:2014wqo,CLAS:2020rue,Schiavilla:2006xx,Sargsian:2004tz,Alvioli:2007zz,Alvioli:2008rw,Hen:2016kwk}.    

Knockout reactions which include detection of recoil nucleons over the full range of momenta are an important tool to measure these SRC pairs and to understand the nuclear ground state.  In addition, they may  also lead to a final resolution of the long standing EMC puzzle
where it has been hypothesized that  SRC pairs and the EMC effect are connected~~\cite{Weinstein:2010rt,Hen:2016kwk,Schmookler:2019nvf,Segarra:2020plg}.  
If the connection is real, then the measured bound-nucleon structure will be modified for events with high-momentum nucleon recoils, but will be unmodified (equal to free nucleon structure) for events with low-momentum recoils.  

The Electron-Ion-Collider (EIC) is an ideal machine for spectator-nucleon tagging measurements, since all recoiling nucleons are boosted in the ion-beam direction to high momentum in the lab frame. This boost makes recoil nucleons  easy to detect with the EIC's far forward detectors~\cite{AbdulKhalek:2021gbh}.  In addition, the EIC will reach much higher momentum transfers than  previous SRC measurements.  The current design of the EIC detectors should allow for a nearly full acceptance for the far-forward going protons, neutrons, and nuclear fragments.  

In this work, we explore the feasibility of tagged spectator-nucleon measurements at the EIC.  We  use the well-known quasi-elastic (QE) knock-out reaction to map out the acceptance for recoil and struck nucleons. 
The recoil nucleon distributions  should be the same for QE and DIS scattering, since in both reactions, the electron scatters from one nucleon in an SRC pair and its partner spectator nucleon recoils.
We will show that such tagged measurements will  be possible at the EIC and that  current double- and triple-coincidence short-range correlation and color transparency measurements can be done over a wide range of momenta and a large range in four-momentum transfer.

\section{Simulation Framework}
\begin{figure}[tb]
\centering
 \includegraphics[trim=0 50 0 0,width=0.49\textwidth]{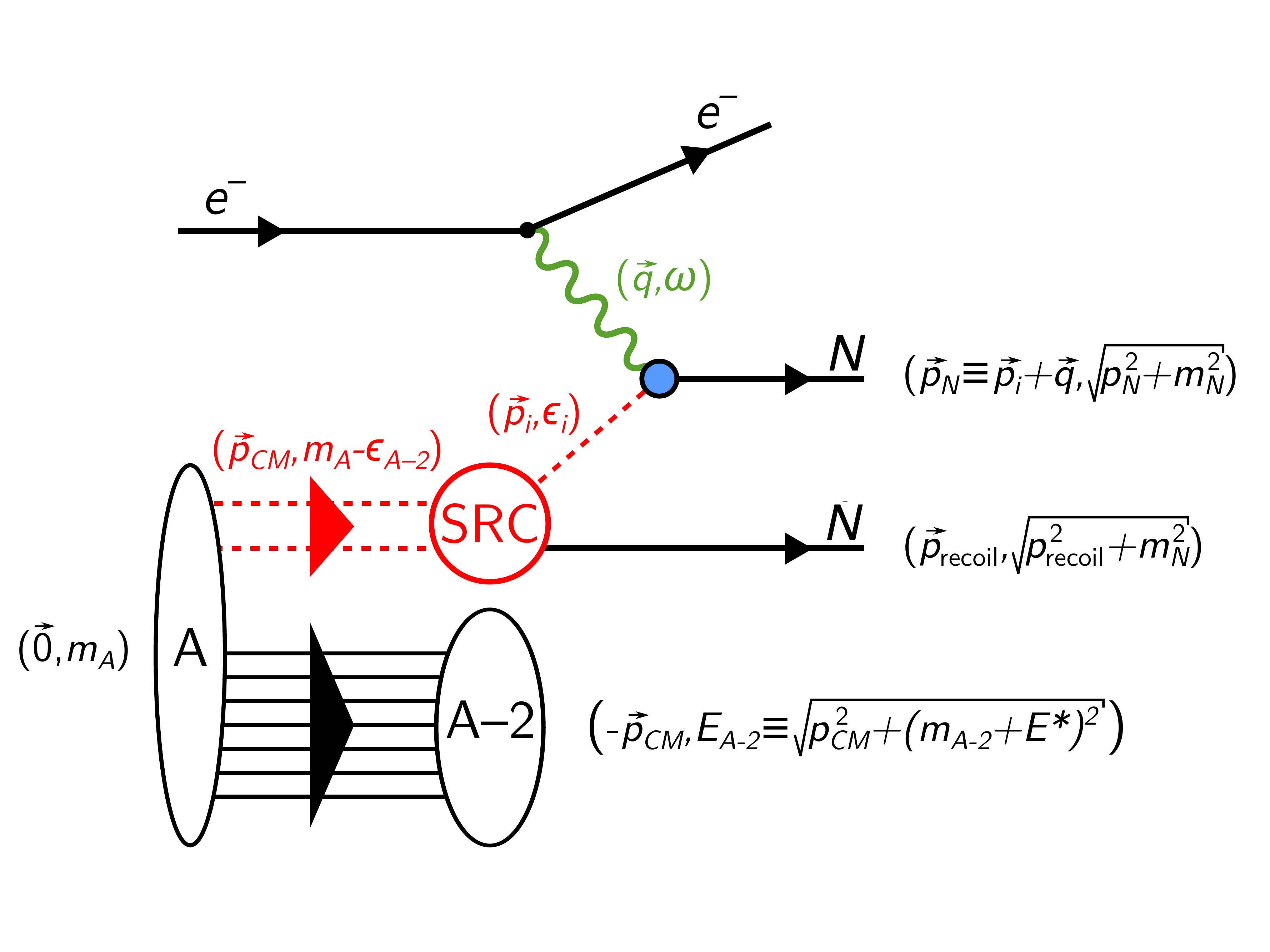}
\caption{Reaction diagram for quasi-elastic (QE) electron scattering off nucleons in a SRC pair. 
The nuclear ground state is assumed to be factorized according to the GCF into an off-shell SRC pair with momentum ${\vec{p}}_{c.m.}$ and a spectator $A - 2$ nuclear system with momentum $-{\vec{p}}_{c.m.}$. 
In the limit of large momentum-transfer, the nucleons in the SRC pair are assumed to further factorize into an active off-shell (lead) nucleon that absorbs the momentum transfer and an on-shell recoil nucleon that is a spectator to the reaction. Figure taken from~\cite{Wright:2021dal}. }
\label{fig:reaction}
\end{figure}

We simulated the reactions using 
 the Generalized-Contact-Formalism (GCF) \cite{Pybus:2020itv,Wright:2021dal,Weiss:2015mba,Weiss:2016obx,Cruz-Torres:2019fum}
to generate QE electron-scattering events from SRC pairs, BeAGLE  (Benchmark $eA$ Generator for LEptoproduction~\cite{beagle}) 
to model nucleon evaporation from the remnant $A-2$ nucleus, and Geant4~\cite{GEANT4} based detector simulations for the EIC (\textit{g4e}~\cite{g4e} and \textit{EICroot}~\cite{EICroot}) to study detector acceptances and resolutions.
We used two energy settings for the electron and ion beams: (1) 5 GeV $e^{-}$ and 41 GeV/A  ions (5x41), and (2) 10 GeV $e^{-}$ and 110 GeV/A ions (10x110).

The diagram for QE scattering from an SRC pair is shown in Fig.~\ref{fig:reaction}.
The event generator assumes that the electron  scatters from a single nucleon via the exchange of a virtual photon with momentum $\vec{q}$ and energy $\omega$, and four-momentum transfer squared $Q^2=\vec q^2 - \omega^2$.
The struck or \textit{leading} nucleon has momentum $\vec{p}_{1}$ and energy $\epsilon_{1}$ before the interaction and momentum $\vec{p}_{1} + \vec{q}$ and energy $\epsilon_{1} + \omega$ after.
The  \textit{recoil} partner nucleon is treated as an on-shell spectator and is emitted with momentum $\vec{p}_{2} = \vec{p}_{c.m.} - \vec{p}_{1}$ and corresponding energy. $\vec{p}_{c.m.}$ is the center of mass motion of the SRC pair. In the GCF generator, the $A-2$ system remains intact with momentum $-\vec{p}_{c.m.}$ and energy $E_{A-2} = \sqrt{\vec{p}_{c.m.}^{2} + (m_{A-2}+E^{*})^{2}}$ where $E^{*}$ is the average excitation energy of the $A-2$ system.

The Generalized-Contact-Formalism (GCF) generator~\cite{Pybus:2020itv} assigns plane-wave impulse approximation (PWIA) cross-section weights for each event. The cross-section is factorized and described by a kinematical factor, the off-shell electron-nucleon cross-section~\cite{DeForest:1983ahx} and the nucleon spectral function. In the GCF, the nucleon spectral function is approximated by a sum over all possible $NN$-SRC pair two-nucleon spectral functions and corresponding contact coefficients (see Ref.~\cite{Pybus:2020itv} for details).  The GCF model successfully describes measured electron and proton scattering data~\cite{Duer:2018sxh,Schmidt:2020kcl,Pybus:2020itv,Korover:2020lqf} as well as ab-initio calculations of one- and two nucleon densities~\cite{Weiss:2016obx,Cruz-Torres:2019fum}.

We used the same GCF model parameters as Ref.~\cite{Cruz-Torres:2019fum}, namely SRC pair relative-momentum distribution calculated with the AV18 $NN$ interaction, and a Gaussian SRC pair c.m. momentum distribution~\cite{CiofidegliAtti:1995qe,Cohen:2018gzh} with a gaussian width (in each Cartesian direction) of $\sigma_{c.m.} = 150$ MeV/c~\cite{Cohen:2018gzh}. 
$E^{*}$ was sampled from a Gaussian distribution with a mean value of $17.3$~MeV/c and a width of $9.6$~MeV/c, based on a study of $J/\psi$ diffractive events using BeAGLE. In that study, intranuclear cascading was turned off and events were selected where the hard collision knocked out two and only two nucleons, leaving an excited $A-2$ nuclear remnant.
These $E^{*}$ values are within the range used in  Refs.~\cite{Schmidt:2020kcl,Weiss:2020bkp}.

The event generation was done in  the rest frame of the target nucleus. The generated events were processed through BeAGLE~\cite{beagle}, run without any rescattering of  the outgoing nucleons, to determine the low-momentum, low-angle "evaporation" nucleons or light nuclei  produced by the nuclear breakup of the $A-2$ nuclear remnant. BeAGLE handles the nuclear remnant and nuclear breakup using FLUKA \cite{fluka}.   In the case of deuterium, no remnant exists and the events are not modified by BeAGLE.

Before the events were passed through \textit{g4e} or \textit{EICroot}, all particles were boosted into the collider frame of the EIC and a crossing angle is applied between the electron and ion beam.  We assumed a crossing angle of 25~mrad consistent with the EIC Yellow Report~\cite{AbdulKhalek:2021gbh}.

The Geant4 simulation program \textit{g4e} implements the detectors according to the reference design of the Yellow Report. We  defined that a particle was accepted if it created a sufficient amount of hits in sub-detectors, thus allowing a reconstruction of its track.
In version 1.3.8 of \textit{g4e}, not all detectors have a detailed sub-structure, e.g., the Zero-Degree-Calorimeter (ZDC). In this case, we assumed a particle was accepted if it crossed the detector volume within its fiducial region.  The events were also processed by \textit{EICroot} to verify and cross-check the results obtained by \textit{g4e}.

\begin{figure}[tb]
\centering
 \includegraphics[width=0.49\textwidth]{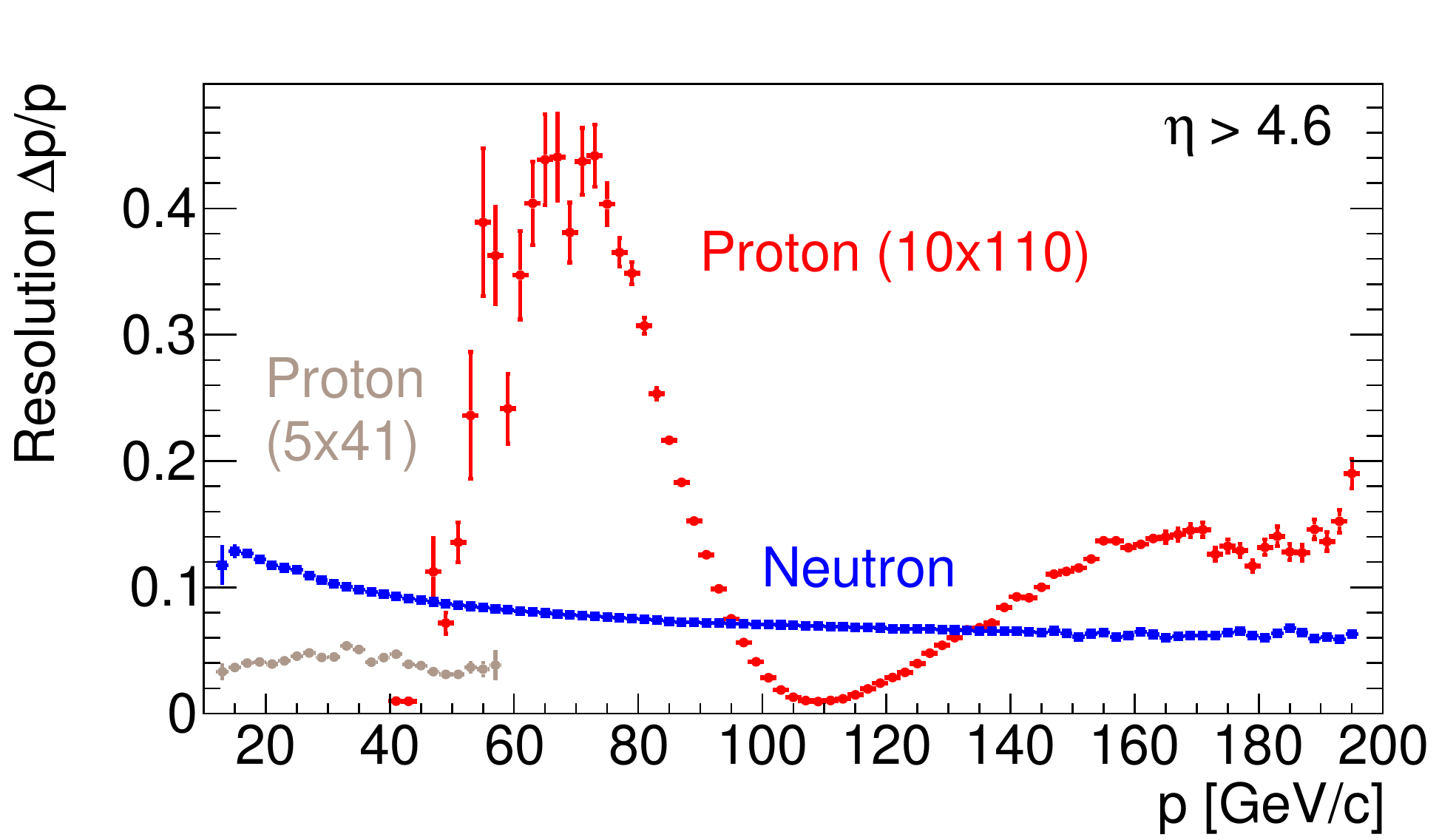}
\caption{Momentum resolution, $\Delta p/p$,  as a function of momentum in the lab frame for neutrons (blue), protons for the 5 GeV x 41 GeV/A beam energy setting (tan) and protons for the 10 GeV x 110/A setting (red) based on previous studies for recoil tagging \cite{Tu:2020ymk} with the Yellow Report detector requirements \cite{AbdulKhalek:2021gbh}.  The particles have a pseudorapidity $\eta > 4.6$.}
\label{fig:smearingmom}
\end{figure}

In the last step of the analysis, the accepted leading and recoil nucleon momenta were smeared based on previous resolution studies \cite{Tu:2020ymk}.
The nucleon momentum smearing is shown in Fig.~\ref{fig:smearingmom}. For neutrons (blue, Fig.~\ref{fig:smearingmom}), the smearing is all based on reconstruction in one detector, namely the ZDC, which has the same energy and angular resolution as described in the Yellow Report, $\frac{\Delta E}{E} = \frac{50\%}{\sqrt{E}} \oplus 5\%$ and $\frac{\Delta\theta}{\theta} = \frac{3\rm{mrad}}{\sqrt{E}}$ \cite{AbdulKhalek:2021gbh}. The protons, however, rely on multiple subsystems for reconstruction. For the 110 GeV/A ion beam case (red, Fig~\ref{fig:smearingmom}), the off-momentum detectors in the far-forward ion direction provide virtually all of the acceptance coverage for the final-state protons, while in the 41 GeV/A case the $B0$ detector covers most of the protons (tan, Fig~\ref{fig:smearingmom}). The resolution function for the off-momentum detectors is dominated by the transfer matrix uncertainty, as described in the Yellow Report \cite{AbdulKhalek:2021gbh}. We also use the Yellow Report values to smear the electron momenta.

The detectors in the far-forward ion side of the EIC detector play an important role to detect the recoil SRC nucleons. The far-forward detectors consist of 
the $B0$ silicon tracker, the Off-Momentum Detectors (OMD), the Roman Pots (RP), and the Zero-Degree Calorimeter (ZDC). They cover angles of 6 to 20\,mrad for the $B0$, 0 to 5\,mrad for the OMD, ~0.5 to 5\,mrad for the Roman Pots, and 0 to ~4\,mrad for the ZDC, respectively. In the ZDC case, the polar angle acceptance is limited by the magnet apertures, but  extends up to $\sim$5\,mrad on one side of the lattice.  Neutrons are detected in the ZDC.
All angles are with respect to the ion beam. 

Particles with angles larger than 30\,mrad can be detected in one of the central detectors. Only protons can be detected in the full far-forward range from ~0 to 20\,mrad, except for the transition region between the RP and B0 detector acceptance.  There is a gap in acceptance between 20\,mrad and 30\,mrad due to the transition between the far-forward and main detector endcap region, where there is no space for detector instrumentation. However, most of our recoil or leading nucleons are scattered to angles outside this gap.

\section{Results}
\begin{figure}[tb]
\includegraphics[trim=0 120 0 50,width=0.48\textwidth]{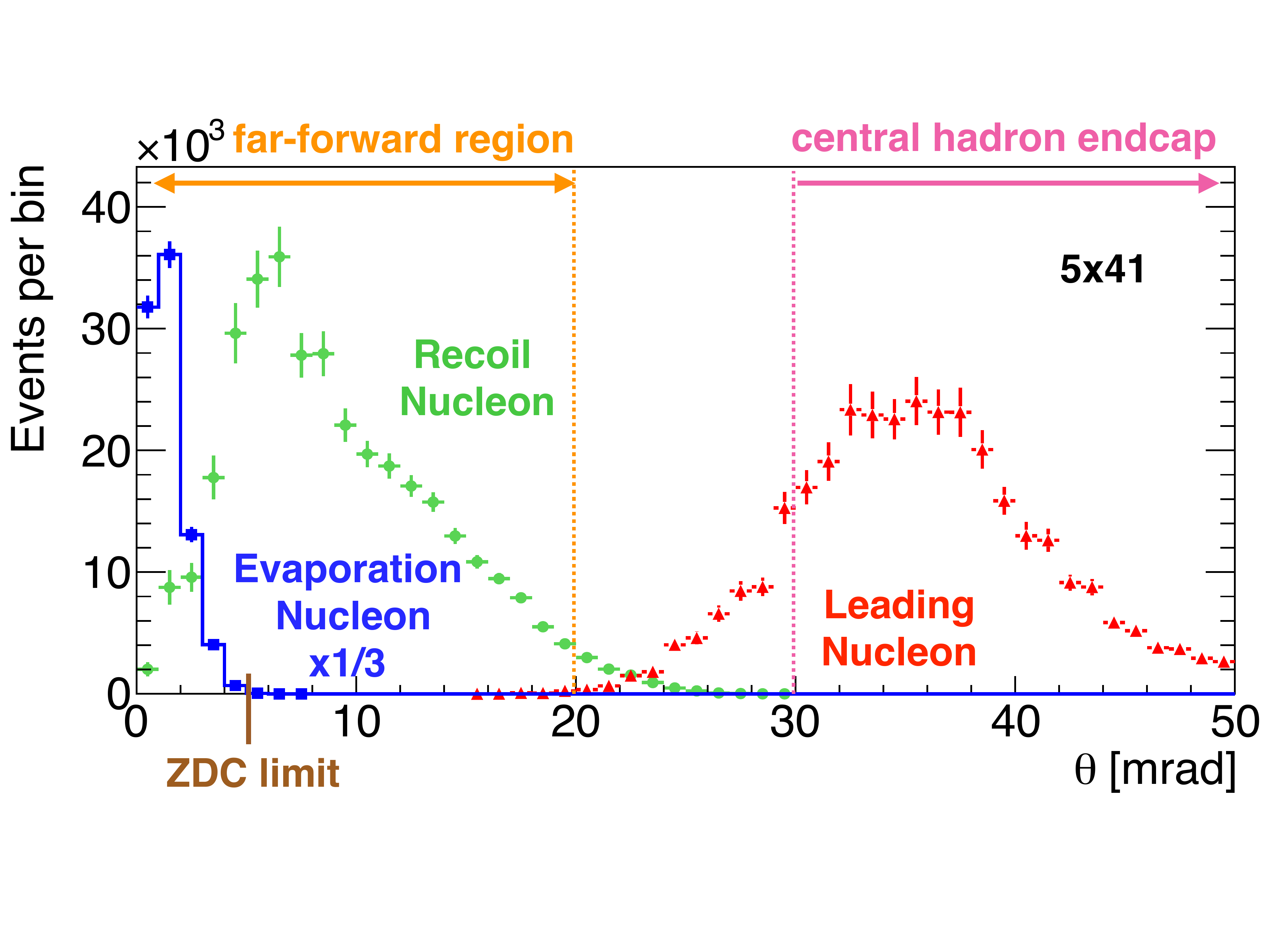} 
\includegraphics[trim=0 120 0 10,width=0.48\textwidth]{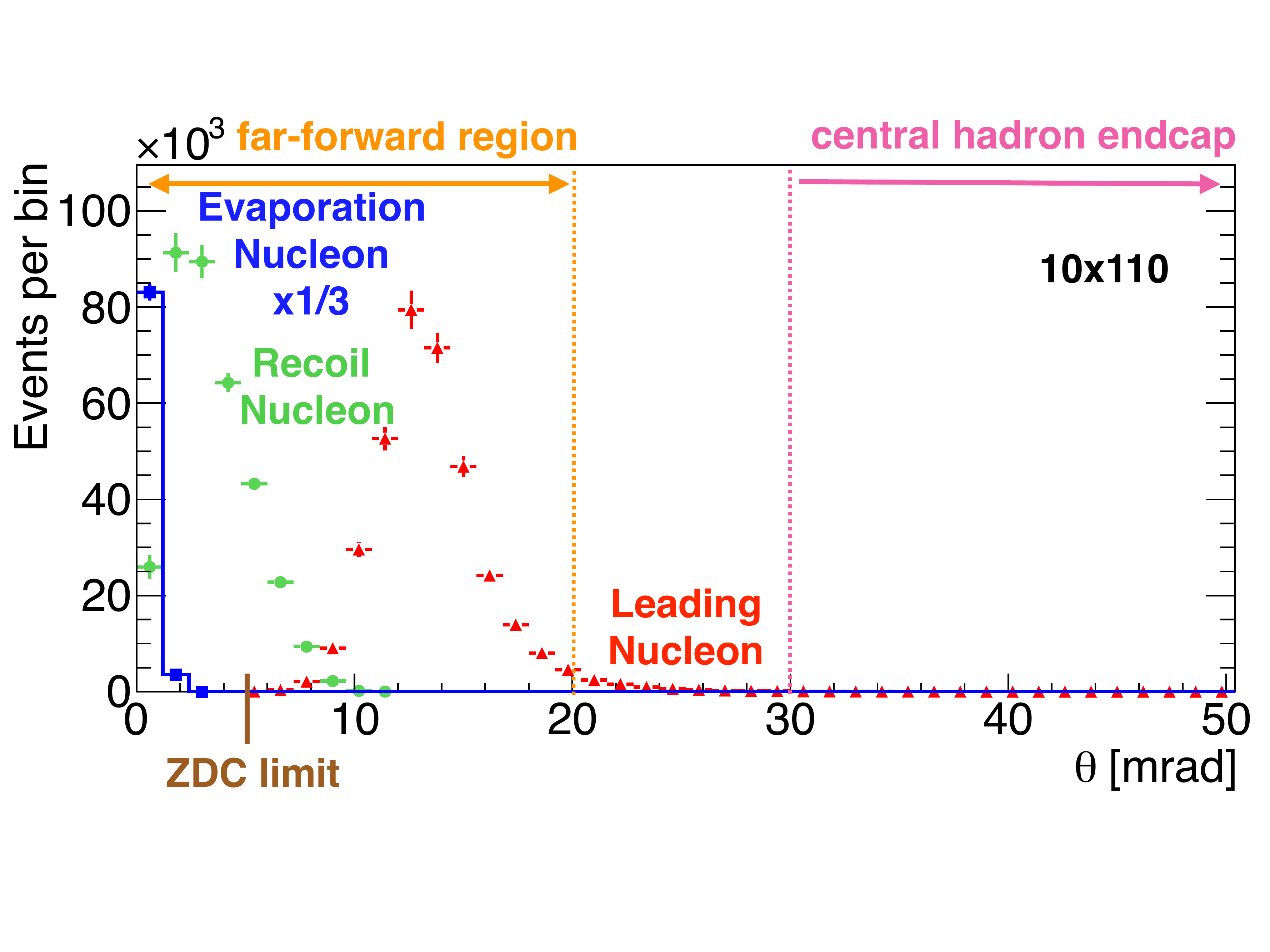}
\caption{Angular distribution of leading (red) and recoil (green) nucleons from the SRC pair as well as the evaporation nucleons (blue) from the A-2 remnant nucleus. This is for $e$C interactions with electron and ion beam energies (in GeV) of  5x41 (top) and 10x110 (bottom) without a crossing angle at the interaction point. This corresponds to the angle with respect to the ion beam if a crossing angle is applied. The orange area marks the acceptance of the far-forward detector and  the magenta region marks the acceptance of the central hadron endcap detector. The brown line marks the limit of the Zero Degree Calorimeter.  The area between 20 and 30 mr  is the transition region between the far-forward detectors and main detector endcap where no detector is installed due to space constraints. The distribution of the evaporation nucleons  has been scaled by 1/3 for better visibility. }
\label{fig:thetadist}
\end{figure}

We studied  QE two-nucleon knockout of SRC pairs in $eD$ and $e$C interactions. Deuterium  can only have the reaction $d(e,e'NN)$ with either a leading proton and recoil neutron $d(e,e'pn)$ or a leading neutron and recoil proton $d(e,e'np)$. Carbon can have all combinations  protons and neutrons and leading and recoil nucleons, C$(e,e'pp)X$, C$(e,e'nn)X$, C$(e,e'np)X$, and C$(e,e'pn)X$.

For both beam energy settings, 5x41 and 10x110, the number of generated events corresponds to an integrated luminosity of $10\,\mathrm{fb}^{-1}$, which is expected in less than one year of EIC operation. The simulations did not include final-state interactions of the outgoing nucleons. 
For the analysis, we applied quasi-elastic SRC selection cuts: $Q^{2} > 3\,\mathrm{GeV}^2$ and $x_{B} = Q^2/2m\omega > 1.2$ based on the generated values.
The  resulting  angular distributions and $Q^{2}$ coverage are similar for both $eD$ and $e$C simulations. Therefore, we  only show the $e$C results.

Fig.~\ref{fig:thetadist} shows the angular distribution of the evaporation, recoil and leading nucleons  for the $e$C simulations with the 5x41 and 10x110 settings. Since no crossing angle is applied for these plots, the angle corresponds to the relative angle between the forward going ion beam and scattered hadrons.
In both settings the recoil and leading nucleons are well separated in angle which allows for a clean distinction between them. The generated distributions are also similar for protons and for neutrons.

In the 5x41 setting (Fig.~\ref{fig:thetadist}, top),  most of the leading protons and some of leading neutrons will be measured in the central detector. Almost all of the recoil protons will be detected in the far-forward detectors and a small fraction of the recoil neutrons (below 4\,mrad) will be detected in the ZDC.
In the 10x110 setting (Fig.~\ref{fig:thetadist}, bottom), the nucleons are at smaller angles due to the larger forward boost from the higher energy ion beam. 
In this case, the recoil and leading nucleon are both within the acceptance of the far-forward detectors. However, the leading neutrons will not be detected because they are outside the limits of the ZDC. Almost all recoil neutrons will be detected. Both recoil and leading protons will be within the acceptance of the far-forward detectors.
Therefore, the choice of the ion momentum affects the total acceptance of the different SRC pairs ($nn$, $pn$, $np$, $pp$) for triple coincidence measurements.

\begin{figure}[tb]
\includegraphics[trim=0 0 0 0,width=0.48\textwidth]{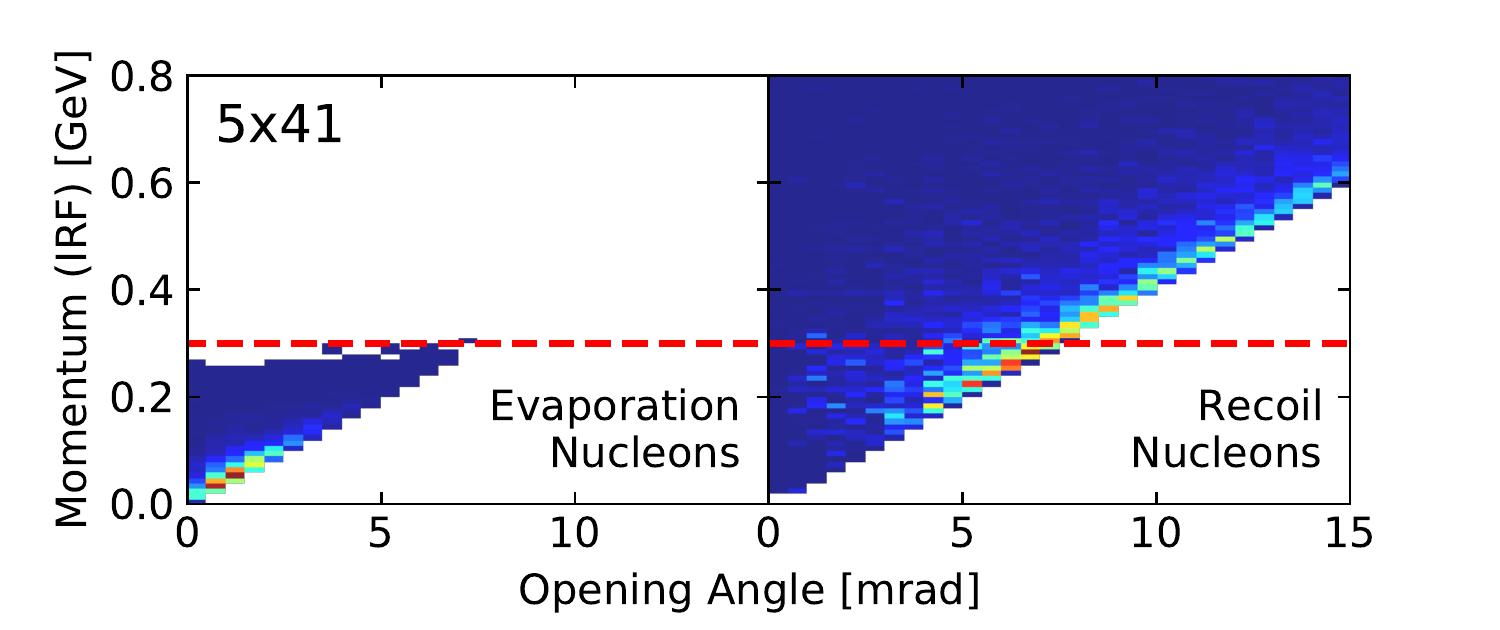} 
\includegraphics[trim=0 0 0 0,width=0.48\textwidth]{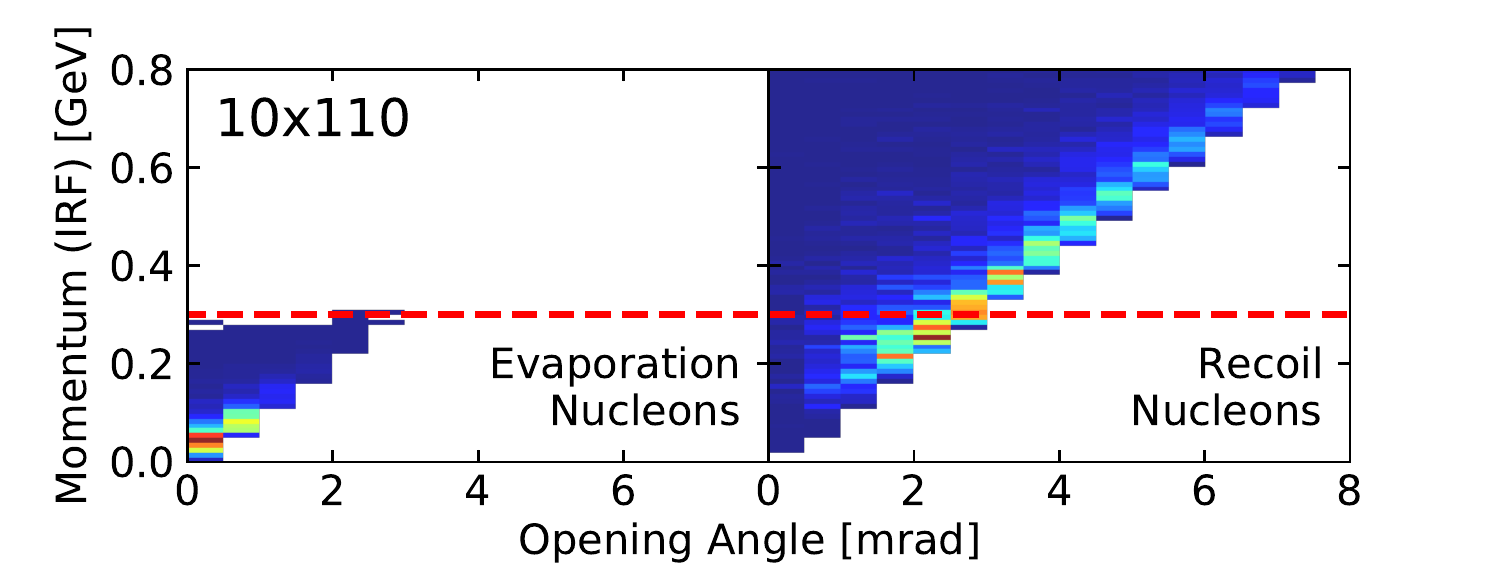}
\caption{Distribution of nucleon momentum in the ion rest frame (IRF) versus angle between nucleon and ion beam direction for evaporation nucleons (left column) and recoil nucleons (right column) without a crossing angle at the interaction point. 
This is for C$(e,e'NN)$X events with electron and ion beam energies (in GeV) of 5x41 (top) and 10x110 (bottom).
The dashed red line marks the applied momentum cut $p_{IRF} > 300\,\mathrm{MeV}/c$ in the analysis which  removes contributions from evaporation nucleons. }
\label{fig:thetamomdist}
\end{figure}

The distributions of the evaporation and recoil nucleons overlap
 at small angles ($\theta < 5\,\textrm{mrad}$) where we cannot separate them.
However, their momentum distributions in the ion rest frame (IRF) are very different. Fig. \ref{fig:thetamomdist} shows the two-dimensional distribution of ion rest frame momentum versus scattering angle  for evaporation nucleons and recoil nucleons. 
There are no evaporation nucleons above $p_{IRF} > 300\,\mathrm{MeV}/c$.
We  cut on this in the further analysis to remove the background contribution from evaporation nucleons. This cut does not affect our physics results, since we are interested in the acceptance of the high momentum SRC-pair recoil nucleons  $p_{IRF} \ge 300\,\mathrm{MeV}/c$.

\begin{figure}[tb]
\centering
 \includegraphics[width=0.50\textwidth]{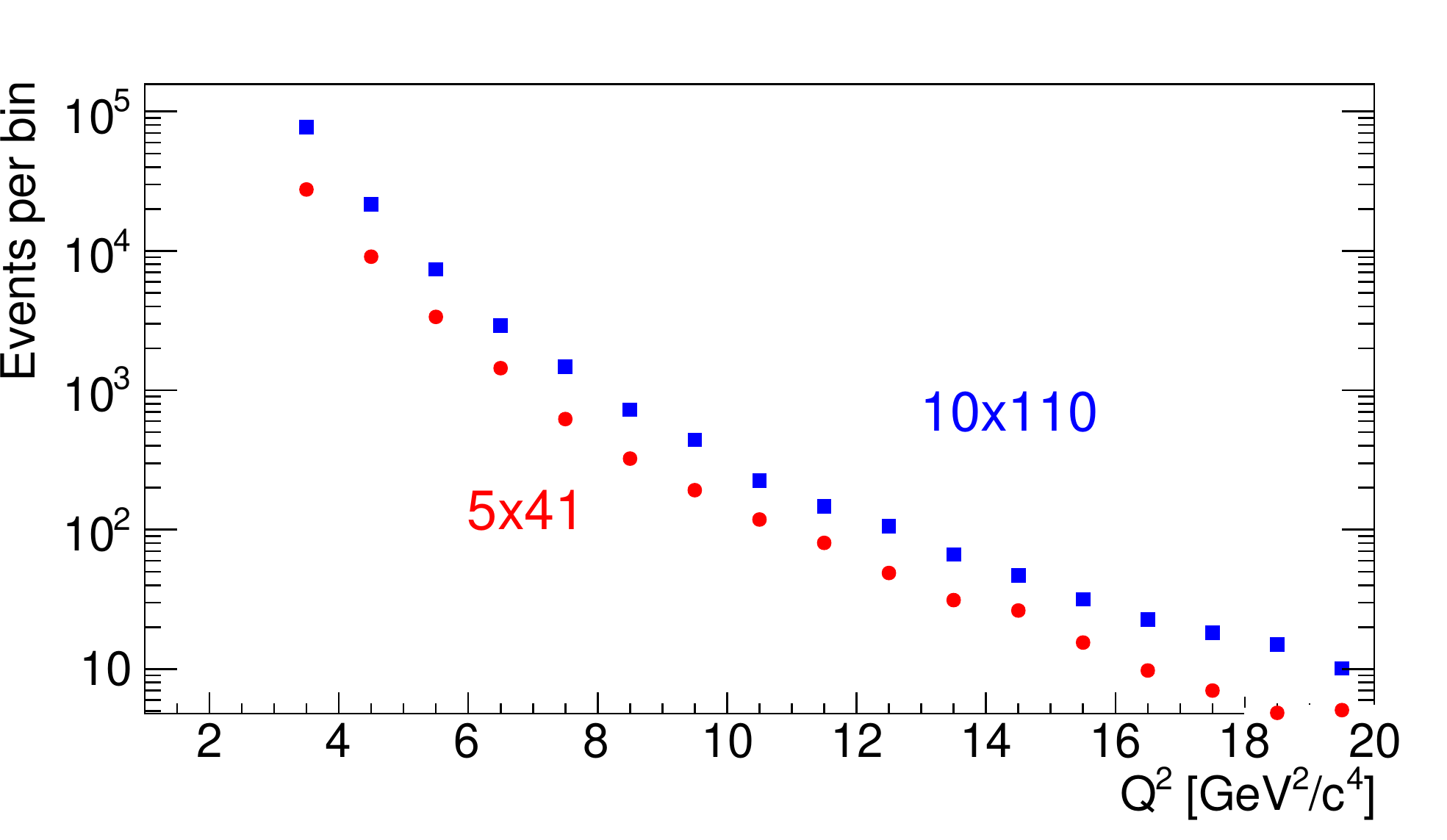}
\caption{Number of simulated events plotted versus $Q^{2}$ for beam settings 10 GeV $e^{-}$ and 110 GeV ions (blue) and 5 GeV $e^{-}$ and 41 GeV ions (red) for $10\,\mathrm{fb}^{-1}$ luminosity where a recoil SRC nucleon is detected in the far-forward detectors for the C$(e,e'NN)$X reaction. Selection cuts $x_B>1.2$, $Q^2 > 3\,\mathrm{GeV}^{2}/c^{4}$ and $p_{IRF} > 300\,\mathrm{MeV}/c$ are applied. An additional scaling factor of 1/2 has been applied to account for transparency corrections.}
\label{fig:q2dist}
\end{figure}

The $Q^{2}$ distribution for events where a recoil nucleon is detected in the
far-forward region is shown in Fig.~\ref{fig:q2dist}  It includes a flat scaling factor of 0.5 to correct for transparency losses of nucleons leaving the Carbon nucleus \cite{Garrow:2001di}. The event rate corresponds to an integrated per nucleus luminosity of $10\,\mathrm{fb}^{-1}$ which equals less than 1 year of EIC operation.  

Reasonable statistics of at least 100 events per bin can be obtained up to $Q^2 \approx 12 \mathrm{GeV}^{2}/c^{4}$ for both energy settings. This extends the reach of previous SRC experiments \cite{Schmidt:2020kcl} by a factor of 3 to 4. The larger $Q^{2}$ reach will allow us to study quasi-elastic two-nucleon knockout for the first time at this large momentum transfer.

\begin{figure*}[tb]
  \includegraphics[trim={0.5cm 0.5cm 0.5cm 0cm},clip,width=0.49\textwidth]{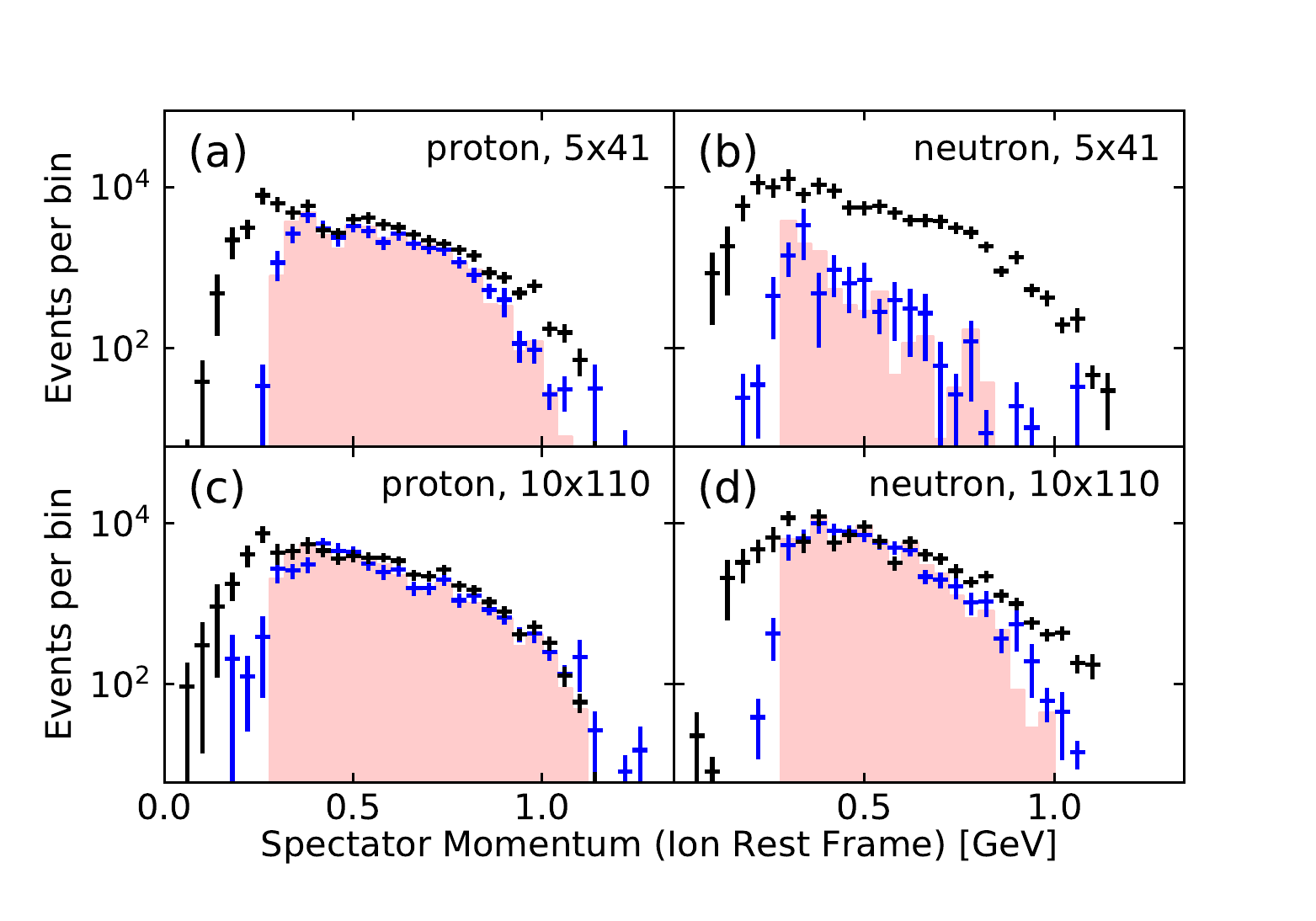}
  \includegraphics[trim={0.5cm 0.5cm 0.5cm 0cm},clip,width=0.49\textwidth]{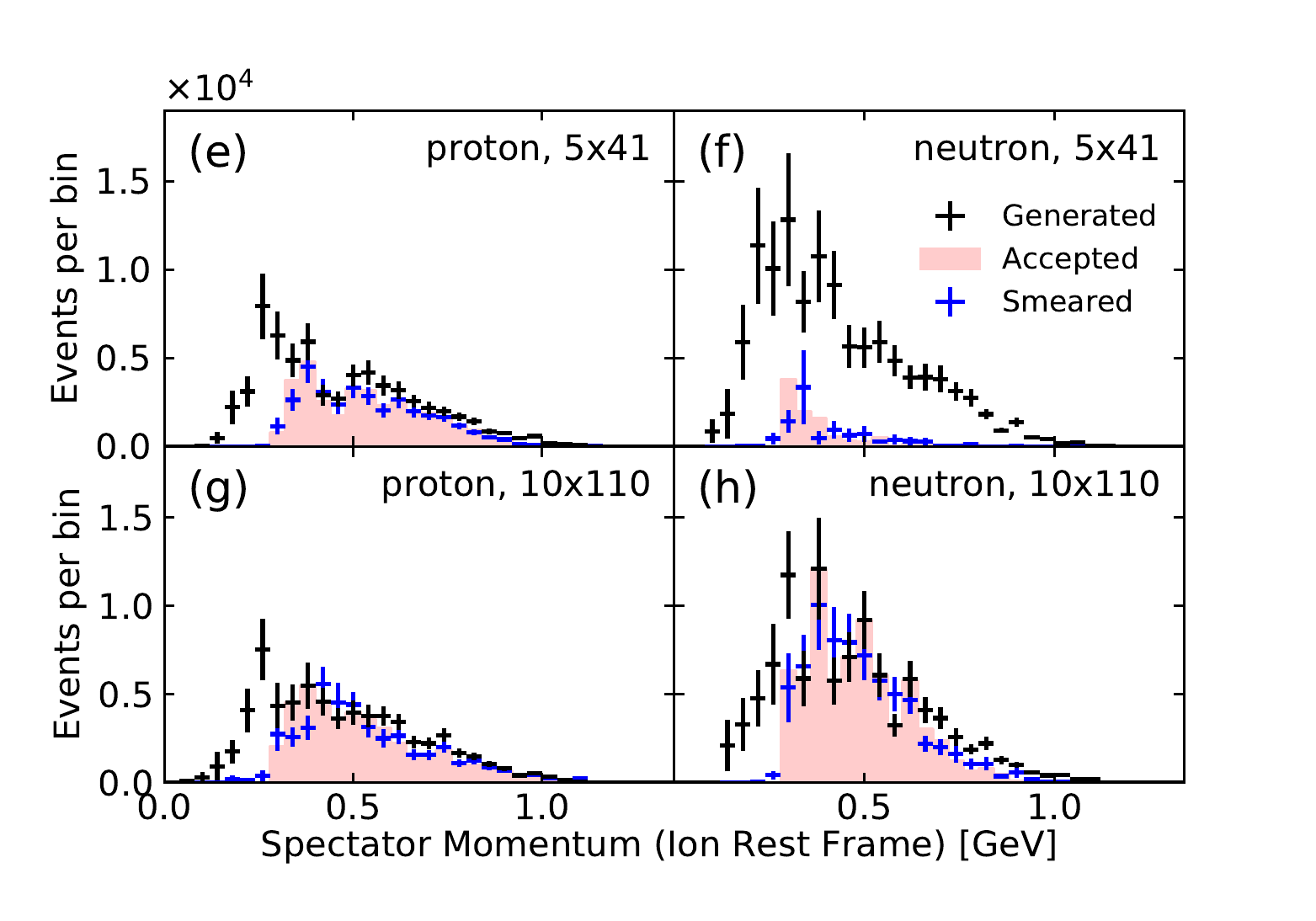}
\caption{Acceptances for tagged recoil nucleons in quasi-elastic SRC breakup in $e$C interactions. Panels (a) - (d) show the distributions in logarithmic scale and (e) - (h) show the same distributions in linear scale. The black points show generated events for $10\,\mathrm{fb}^{-1}$ luminosity and the filled red histograms show the accepted events. The blue points show the distributions after momentum smearing of the accepted events. (Top rows) Results for beam energies 5 GeV $e^{-}$ and 41 GeV ions. (Bottom rows) Results for beam energies 10 GeV $e^{-}$ and 110 GeV ions. Recoil protons are on the left columns in each 4 panel figure and recoil neutrons on the right columns. An additional scaling factor of 1/2 has been applied to all plots to account for nucleon transparency corrections.} 
\label{fig:recoilmomentum_log}
\end{figure*}

The simulated recoil nucleon momentum distributions in the ion rest frame are shown in Fig.~\ref{fig:recoilmomentum_log} for $e$C interactions for neutrons and for protons with the far-forward detector design of the Yellow Report~\cite{AbdulKhalek:2021gbh}.
The recoil proton acceptance is  about 74\% and 90\% 
 for 5x41 and 10x110 settings, respectively. The smaller acceptance for the 5x41 setting stems from the detector gap in the far-forward direction between 20 to 30\,mrad. This mainly influences the acceptance at the largest recoil momenta. The acceptance for both settings is rather similar up to recoil momenta of 900~MeV/$c$ in the ion rest frame.  
The acceptance for recoil neutrons is driven by the Zero-Degree Calorimeter (ZDC). It is about 90\% for 110 GeV/A ions and 30\% for 41 GeV/A ions.
However, recoil momenta above 900~MeV/$c$ are not accepted for 5x41 and barely for 10x110. Below this momentum the acceptance is larger for 110~GeV/A ions. 

We conclude that for recoil nucleons, protons or neutrons, an ion beam energy of 110~GeV/A is preferred to achieve highest acceptance for recoil tagging. However, this setting does not allow for the parallel measurement of leading neutrons due to the limited ZDC acceptance. Only leading protons can be measured in the far-forward region, mostly in the Off-Momentum Detectors. Therefore, a triple coincidence measurement of $nn$ pairs or $np$ pairs with a leading nucleon requires the lower ion beam of 41~GeV/A but with limited acceptance for recoil and leading nucleons. The setting with 110~GeV/A is well suited to detect $pp$ and $pn$ pairs with a leading proton and recoil neutron.
These conclusions are the same for 5 and 10 GeV electron beams. 
The distributions are also not affected by the nucleon momentum smearing.

\begin{figure}[tb]
\centering
 \includegraphics[width=0.49\textwidth]{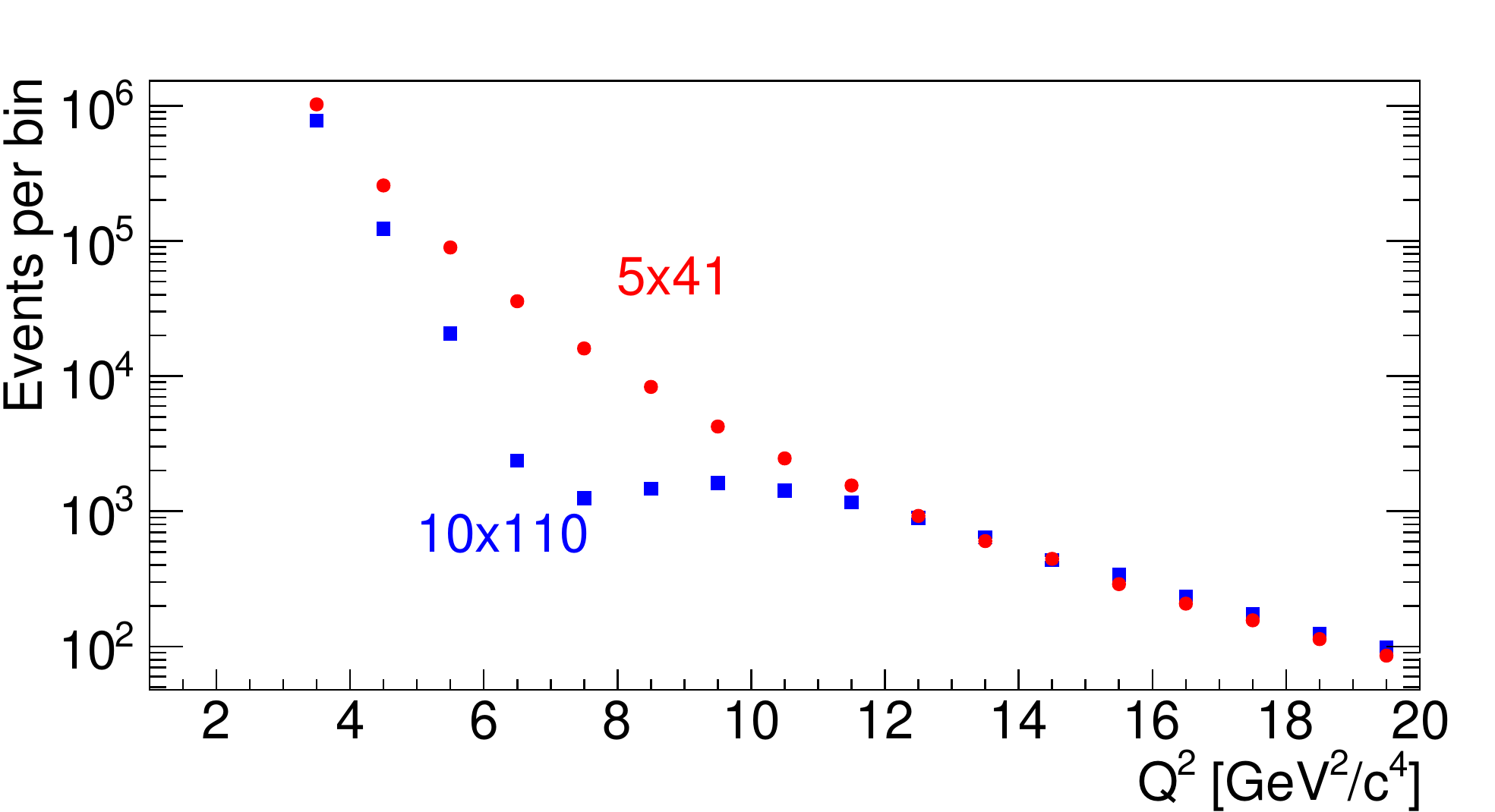}
\caption{$Q^{2}$ distribution for beam settings 10 GeV $e^{-}$ and 110 GeV ions (blue) and 5 GeV $e^{-}$ and 41 GeV ions (red) for $10\,\mathrm{fb}^{-1}$ luminosity for single nucleon knockout where the  nucleon is detected in the far-forward or central detectors. Selection cuts are $Q^2 > 3\,\mathrm{GeV}^{2}/c^{4}$, $p_{miss} < 300\,\mathrm{MeV}/c$ and $E_{miss} < 80\,\mathrm{MeV}/c$. An additional scaling factor of 0.5 has been applied to account for transparency corrections. The dip for the 10x110 setting results from acceptance limitations of the area between the far-forward and central detectors and the limitation of the Zero-Degree Calorimeter. }
\label{fig:q2distMF}
\end{figure}

Besides the two-nucleon knockout simulations, we also performed simulations of C$(e,e'p)X$ single nucleon knockout reactions in the mean field region, i.e. low momentum nucleons in a nucleus. These  are interesting for color transparency measurements \cite{HallC:2020ijh, ONeill:1994znv}.  We used the same beam energy settings as in the case of the SRC simulations. 
We applied typical color transparency cuts on the generated events (see \cite{HallC:2020ijh, ONeill:1994znv}): $p_{miss} < 300\,\mathrm{MeV}/c$, $E_{miss} < 80\,\mathrm{MeV}/c$ and an additional cut $Q^2 > 3\,\mathrm{GeV}^{2}/c^{4}$. Similar to \cite{HallC:2020ijh, ONeill:1994znv}), we define  $p_{miss} = q - p_{lead}$ and $E_{miss} = \omega - T_{lead} - T_{A-1}$ with $T_{X}$ as the kinetic energy of the leading nucleon or $A-1$ nucleus (the momentum of $A-1$ corresponds to $p_{miss}$).

We also required that the electron and knocked-out nucleon  are detected in the far-forward or central detectors. The resulting $Q^2$ distributions for both beam settings are shown in Fig.~\ref{fig:q2distMF} for an
integrated per nucleus luminosity of $10\,\mathrm{fb}^{-1}$.
We expect more than 100 events up to $Q^2\approx 20\,\mathrm{GeV}^{2}/c^{4}$ for both settings. The dip in event-yield for $4 \le Q^2 \le 10\,\mathrm{GeV}^{2}/c^{4}$ for the 10x110  setting  results from two acceptance limitations: (1) Knocked-out nucleons  in the area between the far-forward and central detectors, and (2) Neutrons which are scattered in the far-forward direction outside the ZDC acceptance.

\section{Conclusion}
We  simulated quasi-elastic (QE) two-nucleon knockout of SRC pairs and single-nucleon knockout of mean-field nucleons in $e$C interactions at the future Electron-Ion Collider. 
We studied the reaction at two electron and ion beam settings: 5 GeV $e^{-}$ and 41 GeV/A  ions, and  10 GeV $e^{-}$ and 110 GeV/A ions. 

For two-nucleon knockout reactions,  recoil nucleons  will be detected in the far-forward ion region and leading nucleons will be detected in either the far-forward ion region or the central hadron endcap region. Due to the larger boost from the ion frame to the lab frame, the 110 GeV/A ion beams  maximized the   recoil nucleon acceptance independently of the  electron energy. The acceptance of high momentum recoil neutrons depends on the Zero-Degree Calorimeter in the far-forward region. A much greater fraction of recoil neutrons will be detectable at the higher ion beam energy.
Overall, recoil nucleons with momenta up to 1 GeV/c can be detected with good acceptance and resolution.

Two-nucleon knockout reactions can also achieve a much large range of $Q^{2}$  compared to previous QE SRC experiments \cite{Schmidt:2020kcl}. Reasonable statistics (more than 100 events in one year of running) can be obtained up to $Q^{2}\approx 12 \mathrm{GeV}^{2}/c^{4}$. 
Single-nucleon knockout reactions are feasible up to $Q^2\approx 20\,\mathrm{GeV}^{2}/c^{4}$.

This coverage will enable precision tests of nucleon modification in nuclei by measuring  tagged spectator nucleon DIS as well as expanded searches for color transparency.

\begin{acknowledgments}
This work was funded in part by the US Department of Energy contracts DE-SC0020240, DE-SC0012704, and also DE-AC05-06OR23177, under which Jefferson Science Associates, LLC operates the Thomas Jefferson National Accelerator Facility. We also acknowledge the support of the Jefferson Lab EIC Center. The work of A. Jentsch was further supported by the Program Development program at Brookhaven National Laboratory. The work of J.R. Pybus, was supported by EIC Center fellowships at Jefferson Lab. We thank Barak Schmookler for various discussions and initial studies.
\end{acknowledgments}


%

\end{document}